\documentclass[10pt,letterpaper]{article}
\usepackage{opex3}

\begin{document}

%%%%%%%%%%%%%%%%%% title page information %%%%%%%%%%%%%%%%%%
\title{Observation of noise phase locking in a single-frequency VECSEL}

\author{A. El Amili$^{1}$, V. Pal$^{2}$, F. Goldfarb$^{1}$, R. Ghosh$^{2}$, M. Alouini$^{3,4}$, I.~Sagnes$^{5}$, and F. Bretenaker$^{1}$}

\address{$^{1}$Laboratoire Aim$\acute{e}$ Cotton, CNRS-Universit$\acute{e}$ Paris Sud 11, Campus d'Orsay, 91405 Orsay Cedex, France}
\address{$^{2}$School of Physical Sciences, Jawaharlal Nehru University, New Delhi 110067, India}
\address{$^{3}$Institut de Physique de Rennes, UMR CNRS 6251, Campus de Beaulieu, 35042 Rennes Cedex, France}
\address{$^{4}$Thales Research \& Technology, 1 av. Augustin Fresnel, 91767 Palaiseau Cedex, France}
\address{$^{5}$Laboratoire de Photonique et Nanostructures, CNRS, Route de Nozay, 91460 Marcoussis, France}
\email{Fabien.Bretenaker@u-psud.fr} %% email address is required

% \homepage{http:...} %% author's URL, if desired

%%%%%%%%%%%%%%%%%%% abstract and OCIS codes %%%%%%%%%%%%%%%%
%% [use \begin{abstract*}...\end{abstract*} if exempt from copyright]

\begin{abstract}
We present an experimental observation of phase locking effects in the intensity noise spectrum of a semiconductor laser. These noise correlations are created in the medium by coherent carrier-population oscillations induced by the beatnote between the lasing and non-lasing modes of the laser. This phase locking leads to a modification of the intensity noise profile at around the cavity free-spectral-range value. The noise correlations are evidenced by varying the relative phase shift between the laser mode and the non-lasing adjacent side modes.  
\end{abstract}


%%%%%%%%%%%%%%%%%%%%%%% References %%%%%%%%%%%%%%%%%%%%%%%%%

%%%%%%%%%%%%%%%%%%%%%%%%%%  body  %%%%%%%%%%%%%%%%%%%%%%%%%%
\section{Introduction}

Spontaneously emitted light is well known for having a random phase \cite{Mandel1995}. In particular, the Schawlow-Townes fundamental limit for the phase noise of a laser is due to the phase random walk induced by the photons spontaneously emitted in the lasing mode \cite{GAF}. In the same way, the spontaneously emitted photons, which fall in the non-lasing longitudinal modes, lead to fluctuations of these mode fields around their zero average values \cite{baili2008}.
%which limit the side mode suppression ratio of the laser
This gives rise to an extra intensity noise in the laser at frequencies close to the multiples of the free spectral range (FSR) of the laser. Besides, it has already been shown that the nonlinear effects associated with population pulsations occurring in the laser active medium can lead to some kind of correlations between the phases of the fields present in the non-lasing modes. This has been observed in the case of a gas laser in which the nearest sub-threshold modes on either side of the lasing mode have been shown to be anticorrelated, leading to intensity noise cancellation \cite{harris1992,loudon1994}. 
%it is now well established that a four-wave mixing process due to the nonlinearities of the active medium forces the spontaneous emission to get a certain phase relationship between the longitudinal modes particularly in gas lasers \cite{harris92prl,loudon92}.

More recently, it has been shown that, in the case of a semiconductor laser, the slow-light effect induced by the carrier-population oscillations, associated with the effect of the phase-amplitude coupling, lifts the degeneracy between the beatnote frequencies associated with the two modes neighboring the oscillating mode [6]. This means that, in the case of the vertical external cavity surface emitting laser (VECSEL) of Ref. \cite{karim}, the frequency differences between the lasing mode and the two neighboring modes are no longer equal. Thus, one could expect this deviation from the free spectral range of the laser to forbid any phase locking between the lasing and the two neighboring modes.

The aim of this paper is thus to address the following questions: is there any residual locking (i.e., correlations) between the phases of spontaneously emitted photons falling into sub-threshold longitudinal modes of such a VECSEL? Can we reveal it, and to which extent could such a phase locking mechanism affect the noise spectrum of the laser? In section 2, we present the experimental setup that allows us to measure the relative intensity noise (RIN) spectrum of our single-frequency VECSEL. We report, in section 3, our experimental observations and we give an interpretation of these results. In order to support our interpretation, we detail in section 4 an additional experiment aiming at measuring the phase difference between the two locked side modes.
%But, we have recently shown that because of phase-amplitude coupling (Bogatov effect) and coherent oscillation populations effect (CPO) in a single-frequency semiconductor laser, the cavity FSR has no longer the same value in both side of the lasing mode frequency \cite{karim}. Thus, there is \textit{a priori} no beating of different frequency at each side of the oscillating mode which could prevent the phase-locking (e.g. correlations) between the side modes. We can ask us the following question: is it possible to observe phase-locking (e.g. correlations) between the phases of spontaneously emitted photons falling into sub-threshold longitudinal modes. The aim of this paper is to answer this question and gives experimental evidences in order to explain our observations. In the first section \ref{section-set-up}, we present the experimental setup that allowed us to measure the noise spectrum of our single-frequency laser. We present, in section \ref{section-observation}, our experimental observations and we give an interpretation of these results. In order to support our interpretation, we present in section \ref{section-correlation} a last experience and the corresponding experimental results.

\section{Principle of the experiment and observation of the double-peaked RIN spectrum}\label{section-set-up}

\begin{figure}[]
\centering
\includegraphics[width=0.9 \textwidth]{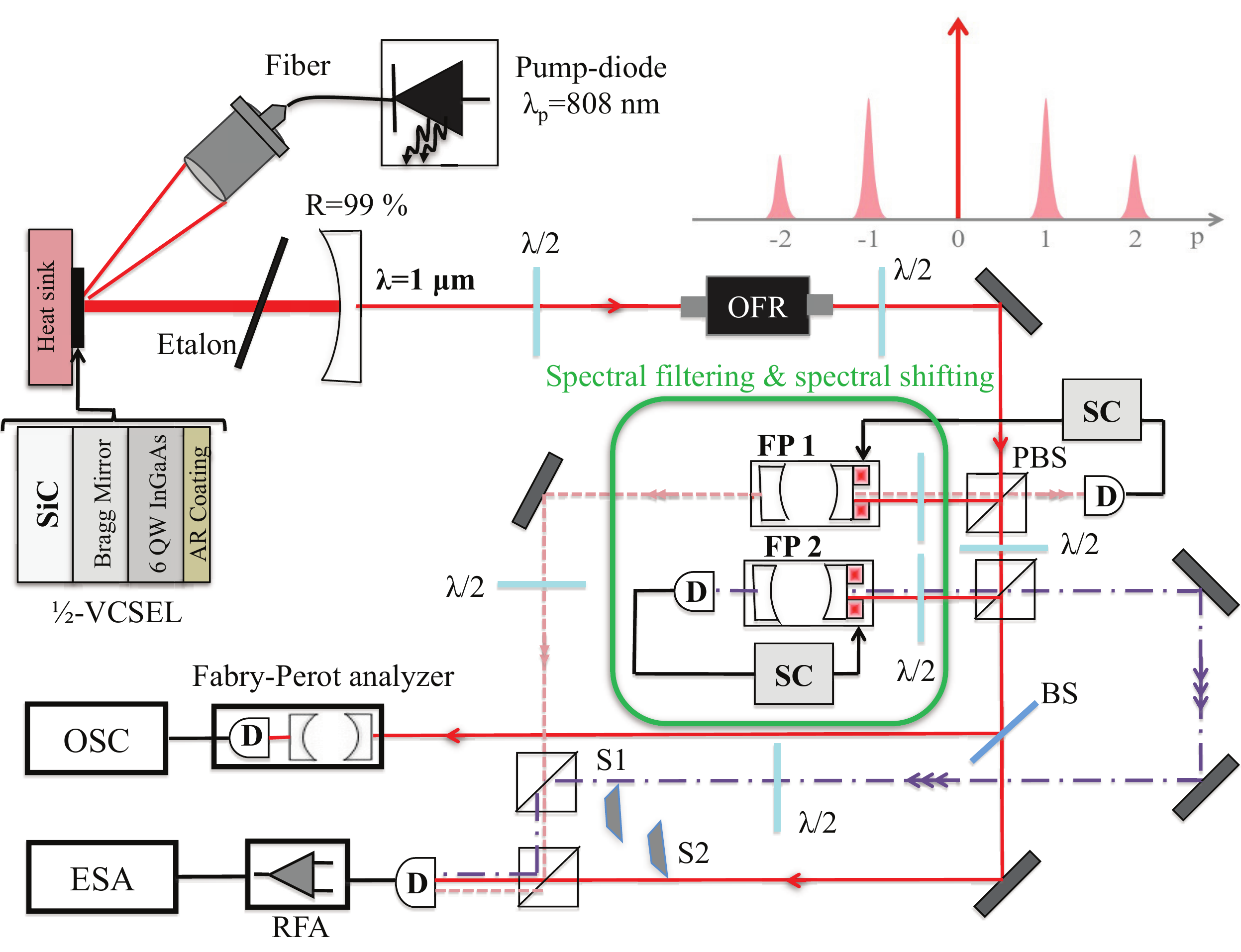}
\caption{Experimental set-up used to measure and analyze the intensity noise spectrum of our VECSEL. OFR: optical isolator based on a Faraday rotator. PBS: polarization beam splitter; BS: beam splitter; SC: Servo-control; D: photodiode; FP1, FP2: Fabry-Perot interferometers; OSC: oscilloscope; S1, S2: shutters used to cut the beam; RFA: radio frequency amplifier; ESA: electronic spectrum analyzer. Inset: Sketch of the laser optical spectrum. $p = 0$ labels the lasing mode while $p \neq 0$ holds for the non-lasing side modes.}\label{set-up}
\end{figure}

The laser used in our experiments is a VECSEL which operates at $\sim1\ \mu$m (see Fig. \ref{set-up}). The planar-concave cavity is formed by a 1/2-VCSEL, which provides the gain and acts as a totally reflecting planar mirror, and an output mirror (10 cm radius of curvature and 99$\%$ reflectivity) separated from the structure by a distance $L$ slightly smaller than 10\ cm. The laser FSR $\Delta$ is then close to 1.5 GHz. In the 1/2-VCSEL, the gain is produced by six InGaAs/GaAsP strained quantum wells grown on a Bragg mirror. This structure exhibits a strong gain anisotropy. Consequently, in the following, all the modes that we consider have the same linear polarizaton. The structure is bonded onto a SiC substrate in order to dissipate the heat towards a Peltier cooler. This multilayered stack is covered by an antireflection coating to prevent any coupled cavity effect. The laser is optically pumped at 808 nm. The gain is broad ($\sim 6\ \mathrm{THz}$ bandwidth) \cite{laurain}. In order to obtain single-frequency operation, a glass \'etalon (200 $\mu$m thick) is inserted inside the cavity. The optical spectrum is continuously analyzed with a Fabry-Perot interferometer to check that the laser remains monomode without any mode hop during data acquisition. The noise spectrum is measured using a 22-GHz bandwidth photodiode and a low-noise amplifier (gain $=$ 47 dB, 20 GHz bandwidth, noise factor $\approx 2.2\ \mathrm{dB}$). The noise spectrum is then analyzed using an electrical spectrum analyzer. The half-wave plates aim at tuning the power incident on the different analyzers.

\begin{figure}[]
\centering
\includegraphics[width=0.9 \textwidth]{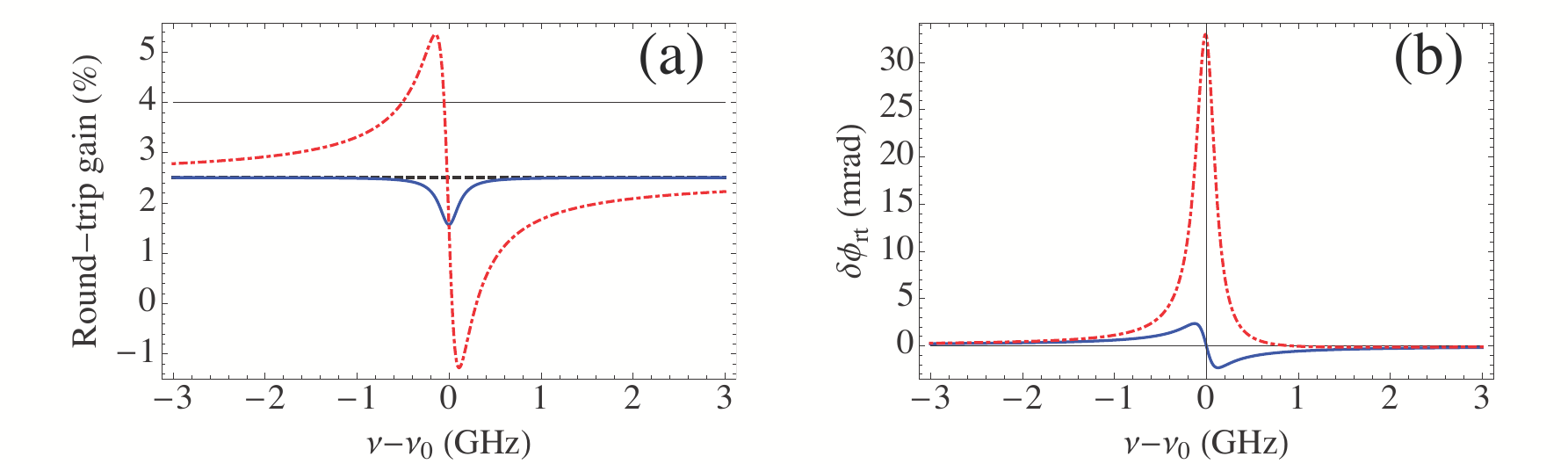}
\caption{(a) Round-trip gain versus side mode frequency detuning $\nu-\nu_0$. The thin line is the unsaturated gain. The dashed line is the saturated gain for the light at $\nu_0$. The full and dotted-dashed lines are the gains seen by the side modes for $\alpha=0$ and $\alpha=7$, respectively. (b) Round-trip phase modification experienced by the side modes for $\alpha=0$ (full line) and $\alpha=7$ (dotted-dashed line).}\label{Fig2}
\end{figure}

Here we focus on the excess intensity noise of the laser in the vicinity of the $p^{\mathrm{th}}$ harmonic of the FSR, i.e., at frequencies close to $\left|p\right|\Delta$. This excess noise is due to the beat notes between the oscillating mode at the frequency $\nu_{0}$ and the spontaneously emitted photons in the adjacent non-lasing modes labeled $\nu_{p}=\nu_{0}+p\Delta$ (see the inset in Fig. \ref{set-up}). Consequently, the excess intensity noise shows up, in the electrical domain, as a peak located at the beat frequency $f_{p}=\left|\nu_{0}-\nu_{p}\right|$. Just above threshold, this excess noise exhibits a Lorentzian shape, whose width is proportional to the extra-losses induced by the intracavity \'etalon on the $p^{\mathrm{th}}$ side mode with respect to the lasing mode \cite{baili2008}. By contrast, at higher pumping rates, the peak shape is no longer purely Lorentzian but is given by the sum of two Lorentzian profiles separated by $\delta f_p=f_{p}-f_{-p}$, which is typically of the order of 100 kHz \cite{karim}. The explanation of the separation of the two peaks holds in the index variations experienced by the non-lasing side modes due to the presence of the lasing mode. Indeed, following Ref.~\cite{karim}, this frequency difference is given by
\begin{equation}
\delta f_p\approx \nu_{0}\frac{L_{m}}{L+n_{0}L_{m}}(\delta n_p + \delta n_{-p}) , \label{eqdeltaf}
\end{equation}
where $n_{0}$ is the bulk refractive index of the semiconductor structure which has a length $L_m$. $\delta n_{\pm p}$ are the modifications of the refractive index of the structure experienced by the $\pm p$ side modes and induced by the dispersion associated with the coherent population oscillation (CPO) effect created by the beat note between the lasing mode and the side modes. In a semiconductor active medium, thanks to the Bogatov effect \cite{Bogatov}, the dispersion is no longer an odd function of the frequency detuning with respect to $\nu_0$. Thus, $\delta n_p \neq -\delta n_{-p}$ and the two beat note frequencies $f_p$ and $f_{-p}$ corresponding to the $p$ and $-p$ modes occur at slightly different frequencies. This is illustrated in Fig.~\ref{Fig2} which displays the gain and dispersion curves experienced by a side mode of frequency $\nu$ in the presence of the lasing mode at frequency $\nu_0$ versus $\nu-\nu_0$ for a Henry factor $\alpha$ equal to 0 (full line) or different from 0 (dotted-dashed line). These plots are obtained using the following expressions for the gain $g\left(\nu\right)$ and the refractive index variation $\delta n\left(\nu\right) = n\left(\nu\right)-n_0$ seen by the side modes \cite{Agrawal2}:
\begin{eqnarray}
g(\nu) &=& \frac{g_{0}}{1+\mathcal{S}}\left\{1-\frac{\mathcal{S}\left[(1+\mathcal{S})+\alpha 2\pi\left(\nu_{0}-\nu\right)\tau_{\mathrm{c}}\right]}{(1+\mathcal{S})^2+\left[2\pi\left(\nu_{0}-\nu\right)  \tau_{\mathrm{c}}\right]^2}\right\},\qquad \label{eqn-gain} \\
 \delta n\left(\nu\right)
 &=& \frac{c}{4\pi\nu_{0}}\frac{g_{0}\mathcal{S}}{1+\mathcal{S}}
\frac{2\pi\left(\nu_{0}-\nu\right)\tau_{\mathrm{c}}+\alpha(1+\mathcal{S})}{\left(1+\mathcal{S}\right)^2+\left[2\pi\left(\nu_{0}-\nu\right)\tau_{\mathrm{c}}\right]^2}\, ,\label{eqn-index}
\end{eqnarray}
with $\tau_{\mathrm{c}}=2\ \mathrm{ns}$, $\mathcal{S}=0.65$, and round-trip gain $G_0=2g_0 L_m=0.04$. This corresponds to the parameters used below.
Here $g_{0}$ is the unsaturated gain and $\mathcal{S}$ the saturation parameter. $\tau_{\mathrm{c}}$ is the lifetime of the carriers and $c$ the speed of light in vacuum. Figure \ref{Fig2} clearly shows that a non-vanishing value for $\alpha$ is mandatory to restrain the saturated dispersion curves from being odd and thus to explain the difference between $f_p$ and $f_{-p}$.

\section{Experimental evidence of the phase locking of the noisy side modes}
\label{section-observation}

\begin{figure}[]
\centering
\includegraphics[width=0.9 \textwidth]{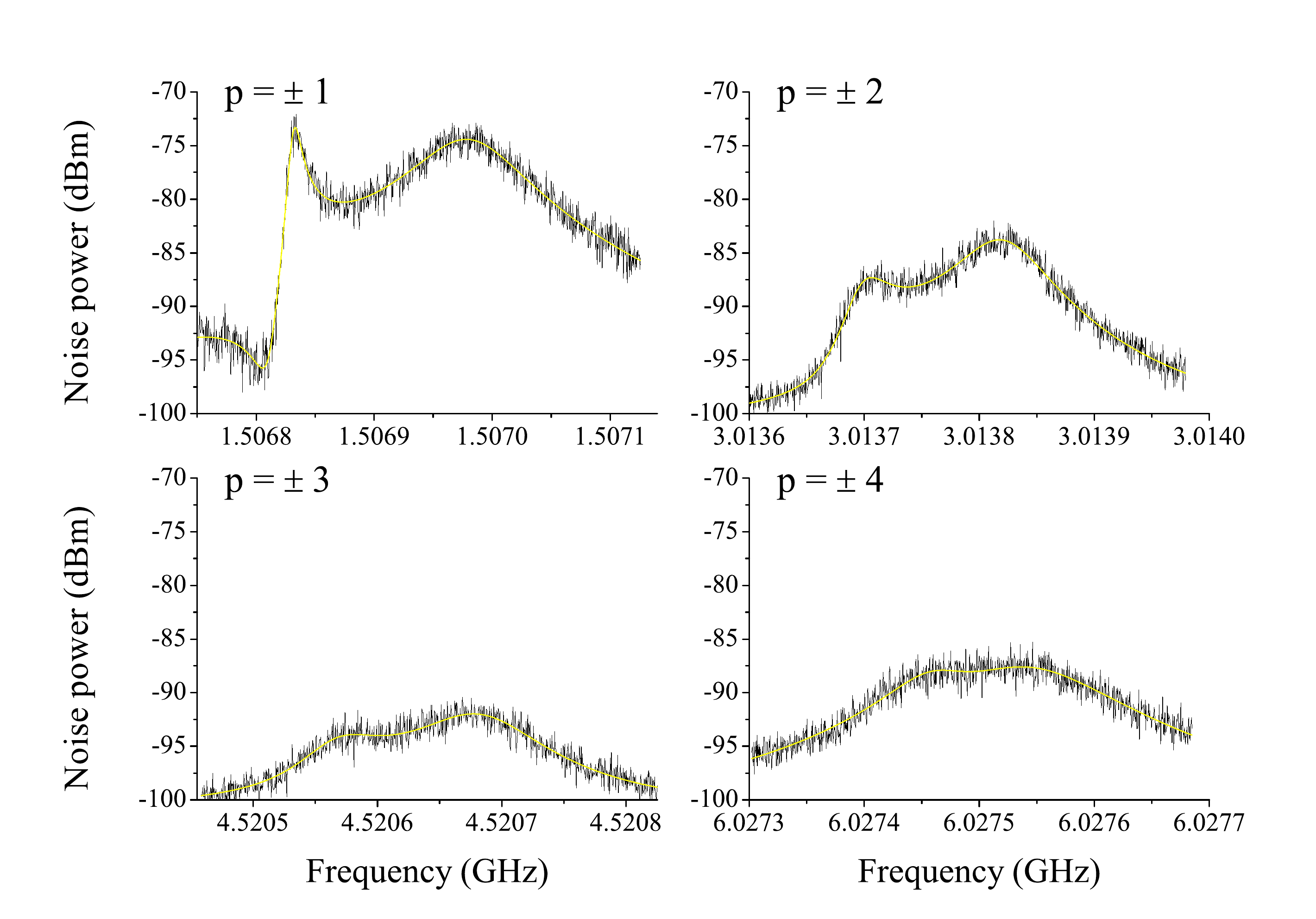}
\caption{Intensity noise spectra at frequencies close to the successive harmonics of the resonator FSR $\left|p\right|\Delta$. The beat note at $\Delta$ (upper left spectrum) shows an intensity noise profile completely different from the spectra taken at higher beat frequencies. The spectrum at $\Delta\approx 1.5\ \mathrm{GHz}$ ($p=\pm1$) is fitted by a coherent sum of Lorentzian profile by using the expression of  Eq. (\ref{eq5}). By contrast, at higher beat frequencies, the spectra for $p=\pm 2,\pm 3,\pm 4$ are fitted by a simple incoherent sum of two Lorentzian profiles as in Eq. (\ref{eq4}). For these spectra, the laser output power is 20~mW.}
\label{Fig3}
\end{figure}
%(Fig. \ref{spectres-harmoniques})

By further increasing the pumping rate, the gain saturation can increase up to a level for which the population oscillations can induce a phase locking between the non-lasing noisy side modes. This is evidenced in Fig.\ \ref{Fig3}. This figure reproduces experimental noise spectra obtained in the same experimental conditions for frequencies close to successive orders $p$ of the laser FSR $\Delta$ ($\Delta\approx 1.5\ \mathrm{GHz}$) with $p=1, \ldots, 4$. As expected from  Eqs.\ (\ref{eqdeltaf}) and (\ref{eqn-index}), each spectrum consists of two peaks separated by a frequency difference $\delta f_p$. However, one can see that the first spectrum, which corresponds to the first side modes ($p=\pm1$), is qualitatively different from the three other ones ($p=\pm2,\pm3,\pm4$). Unlike the three latter ones, this former spectrum cannot be interpreted as the incoherent sum of two Lorentzian peaks. Indeed, the decrease of the noise power at frequencies slightly larger than $1.5068\ \mathrm{GHz}$, followed by a sharp increase and a peak in the spectrum, evidences the existence of a precise phase relation between the two noise spectra corresponding to $p=+1$ and $p=-1$. The existence of this phase relation is definitely proved when one wants to fit the spectra of Fig.\ \ref{Fig3} with an incoherent sum of two Lorentzians, such as:
\begin{equation}
 y = \left|\frac{A_{-1}\gamma_{-1}}{2\mathrm{i}\times2\pi(f-f_{-1})+\gamma_{-1}}\right|^2+ \left|\frac{A_{1}\gamma_{1}}{2\mathrm{i}\times2\pi(f-f_{1})+\gamma_{1}}\right|^2 , \label{eq4}
\end{equation}
where the two Lorentzians centered at $f_1$ and $f_{-1}$ have widths equal to $\gamma_1/2\pi$ and $\gamma_{-1}/2\pi$, respectively. These widths hold for the extra losses imposed by the \'etalon to the non-lasing modes (and also the modification of their gains due to CPO, as explained at the end of the present section) with respect to the lasing mode. Eq.\ (\ref{eq4}) gives a perfect fit for the spectra of Fig.\ \ref{Fig3} corresponding to $p=\pm2,\pm3,\pm4$, but is unable to provide a good fit for the spectrum corresponding to $p=\pm1$. This latter spectrum can be well fitted using a coherent sum of the two Lorentzians, namely:
\begin{equation}
 y = \left|\frac{A_{-1}\gamma_{-1}}{2\mathrm{i}\times2\pi(f-f_{-1})+\gamma_{-1}}+ \frac{A_{1}\gamma_{1}e^{-\mathrm{i}\Phi}}{2\mathrm{i}\times2\pi(f-f_{1})+\gamma_{1}}\right|^2 . \label{eq5}
\end{equation}
In this equation, the phase shift $\Phi$ between the lasing mode and the side modes is expressed by:
\begin{equation}
\Phi = 2\phi_{0}-\left(\phi_{1}+\phi_{-1}\right), \label{eq6}
\end{equation}
where $\phi_p$ is the phase of the mode labeled by $p$. In the case of the first spectrum of Fig.\ \ref{Fig3}, the fit leads to $\Phi=3.5\ \mathrm{rad}$. The fact that  Eq.\ (\ref{eq5}) provides a nice fit of the spectrum corresponding to $p=\pm1$ is a direct signature of phase correlations between the beat frequencies of the lasing mode $p =$ 0 and the spontaneously emitted photons falling into the immediate adjacent modes $p =\pm 1$. 

\begin{figure}[]
\centering
\includegraphics[width=0.6\textwidth]{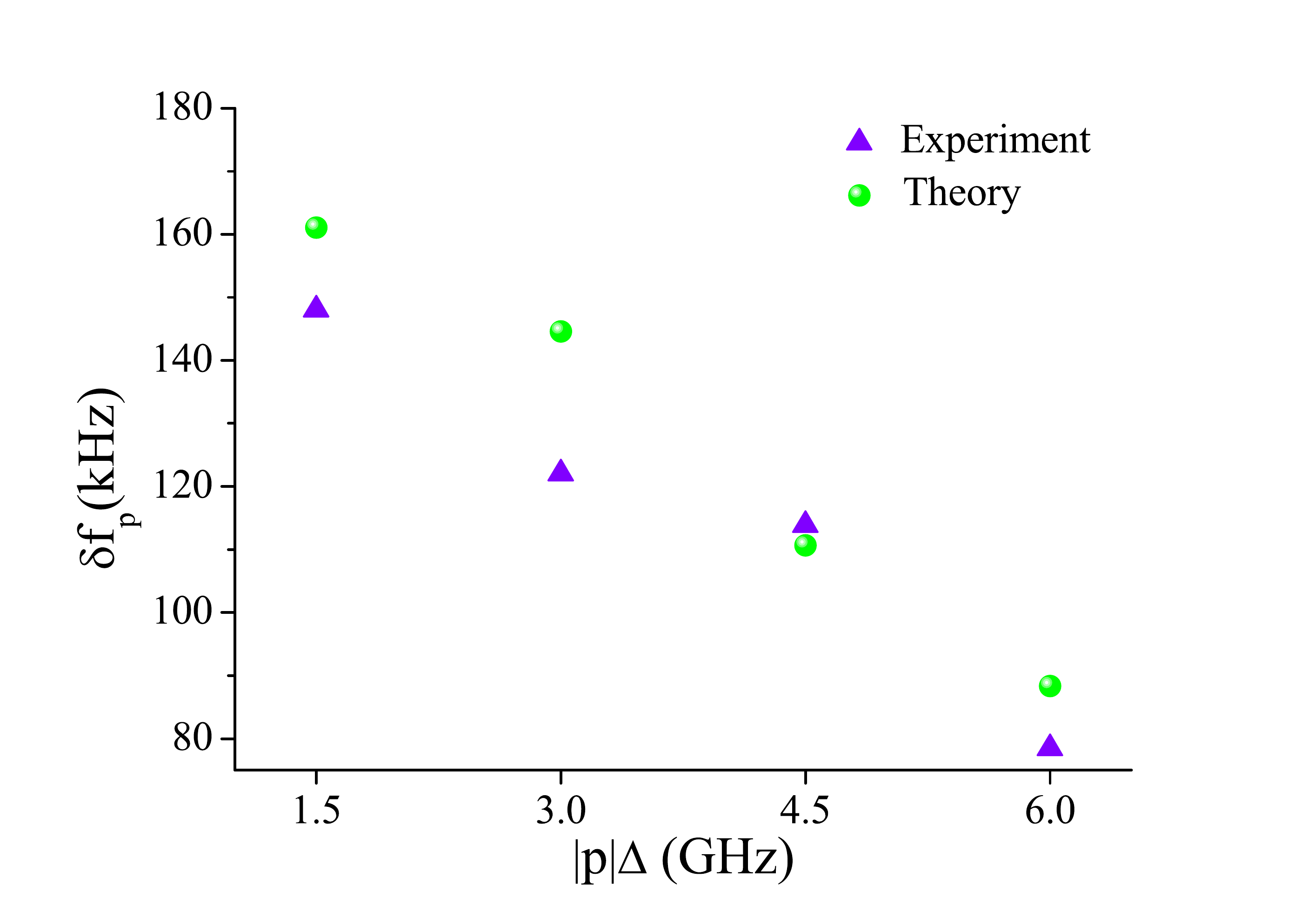}
\caption{Theoretical and experimental evolutions of the frequency shift $\delta f_p$ as a function of the frequency shift $|p|\Delta$ of the side modes. The theoretical evolution is calculated with the same parameters as in Fig.\ \ref{Fig2}. The experimental results vary by about 20~\% from acquisition to acquisition, due to changes in the operation point of the laser.} \label{df-vs-p}
\end{figure}

The fact that the phase correlations become blurred at higher harmonics of the FSR is due to the decrease of the amplitude of the population pulsations at high frequencies. This is due to the finite response time $\tau_{\mathrm{c}}$ of the carriers (of the order of 2\ ns) which limits the bandwidth of the phenomenon. This also explains why the separation $\delta f_p$ between the peaks decreases when the beat frequency increases: the index modifications seen by the two modes are then smaller (see Fig.\ \ref{Fig2}(b)). This is illustrated in Fig.\ \ref{df-vs-p}. In this figure, the experimental values of $\delta f_p$, extracted by fitting the spectra of Fig.\ \ref{Fig3} with  Eqs.\ (\ref{eq4}) and (\ref{eq5}), are compared with the theoretical values obtained using  Eqs.\ (\ref{eqdeltaf}) and (\ref{eqn-index}). A good agreement is obtained, comforting the model presented in reference \cite{karim}.

\begin{figure}[h]
\centering
\includegraphics[width=0.9\textwidth]{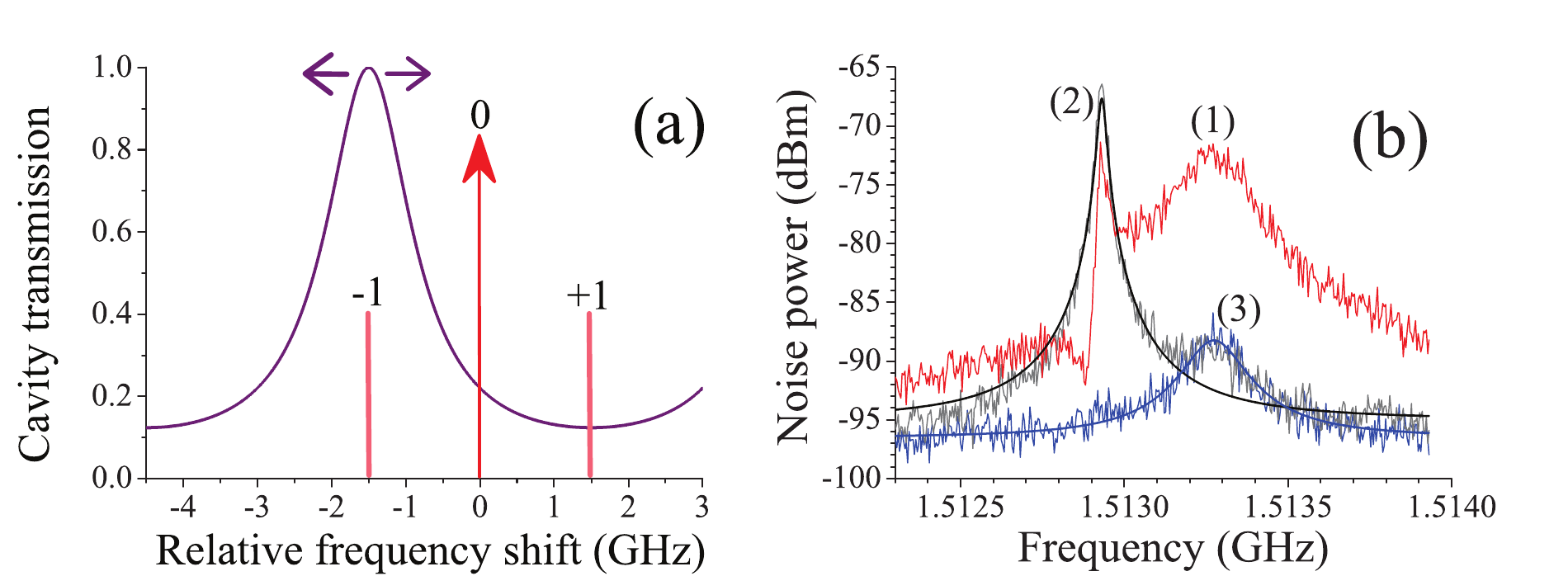}
\caption{(a) Principle of the filtering of the side modes using the transmission of a low-finesse Fabry-Perot cavity. (b) Noise spectrum after transmission of the laser beam by the Fabryt-Perot cavity FP1. The spectra labeled (1), (2), and (3) are obtained with the cavity resonance frequency tuned on the frequencies of modes $p=0$, $-1$, and $+1$, respectively. The spectra labeled (2) and (3) are fitted by Lorentzians. The fit of the spectrum labeled (1) leads to $\Phi=2.75\ \mathrm{rad}$.}
\label{Fig5}
\end{figure}

From now on and for the rest of the present paper, we focus on the first double peak spectrum located near 1.5\ GHz and corresponding to the first side modes labeled $p=\pm 1$. We start by checking the fact that the two peaks which constitute this spectrum can each be attributed to one of the two non-lasing side modes. With this aim in view, we specially designed a Fabry-Perot cavity capable of filtering out one of these side modes. We choose a cavity with a rather poor finesse (about 4) with an FSR of 6\ GHz. Then, at the output of the VECSEL, a part of the beam is injected into this Fabry-Perot cavity (labeled FP1 in Fig.\ \ref{set-up}). The cavity length is stabilized and controlled using a servo loop which maintains the transmission of the lasing mode to a given offset level. By choosing this offset level, we are able to adjust the frequency shift between the resonance frequency of the cavity and the lasing mode frequency, allowing one of the side modes to be perfectly transmitted by the cavity while the other one is almost perfectly reflected. For example, Fig.\ \ref{Fig5}(a) represents the situation in which the cavity is tuned in order to eliminate mode $p=+1$ while maintaining the beatnote between the lasing mode $p=0$ and mode $p=-1$ on the detector which monitors the intensity transmitted by the cavity.

Figure \ref{Fig5}(b) reproduces the experimental spectra obtained when the cavity resonance is centered on the lasing mode (spectrum labeled (1)), on the low-frequency side mode $p=-1$ (spectrum labeled (2)), and on the right frequency side mode $p=1$ (spectrum labeled (3)). The spectrum labeled (1) cannot be fitted as an incoherent sum of the individual spectra labeled (2) and (3) but must be described using eq.\ (\ref{eq5}). This shows that the initial spectrum is indeed composed of two frequency-shifted Lorentzian noise spectra, with a deterministic phase difference. 

The existence of two different beat frequencies $f_{-1}$ and $f_{+1}$ for the two beat notes is a clear evidence of the role of the CPO-induced dispersion spectra of Fig.\ \ref{Fig2}(b). The role of this slow light effect, which is visible in semiconductors thanks to the non-vanishing value of the Henry factor $\alpha$ \cite{karim}, is of course absent in the analysis of Harris, Loudon, and coworkers \cite{harris1992,loudon1994}, who were dealing with a gas laser. Notice also that spectra similar to the one labeled (1) in Fig.\ \ref{Fig5} were also reported earlier \cite{VanExter1994} in the case of an external cavity diode laser, but analyzed using the approach of Ref.\ \cite{loudon1994} which did not take the existence of $\delta f_1$ into account.

Finally, the experiment reported in Fig.\ \ref{Fig5} shows that the peak which has the lower beat frequency corresponds to the beatnote between the lasing mode and the lower-frequency side mode ($p=-1$). The fact that this peak is the narrower one is consistent with the fact that CPO induces an excess gain for this mode (see the red dotted-dashed curve of Fig.\ \ref{Fig2}(a) for $\nu-\nu_0=-1.5\ \mathrm{GHz}$), contrary to the high-frequency side mode $p=1$ which experiences a decrease of its gain due to CPO (see the red dotted-dashed curve of Fig.\ \ref{Fig2}(a) for $\nu-\nu_0=+1.5\ \mathrm{GHz}$). Notice also that the fits of the phase-locked double peak structures lead to values of $\Phi$ which are often close to $\pi$, which is consistent with the fact that mode competition tends to put the two oscillations in antiphase, as evidenced by Loudon \emph{et al.} in their model \cite{loudon1994}.

\section{Role of the phase shift between the lasing and non-lasing modes}
\label{section-correlation}

\begin{figure}[h]
\centering
\includegraphics[width=1.0\textwidth]{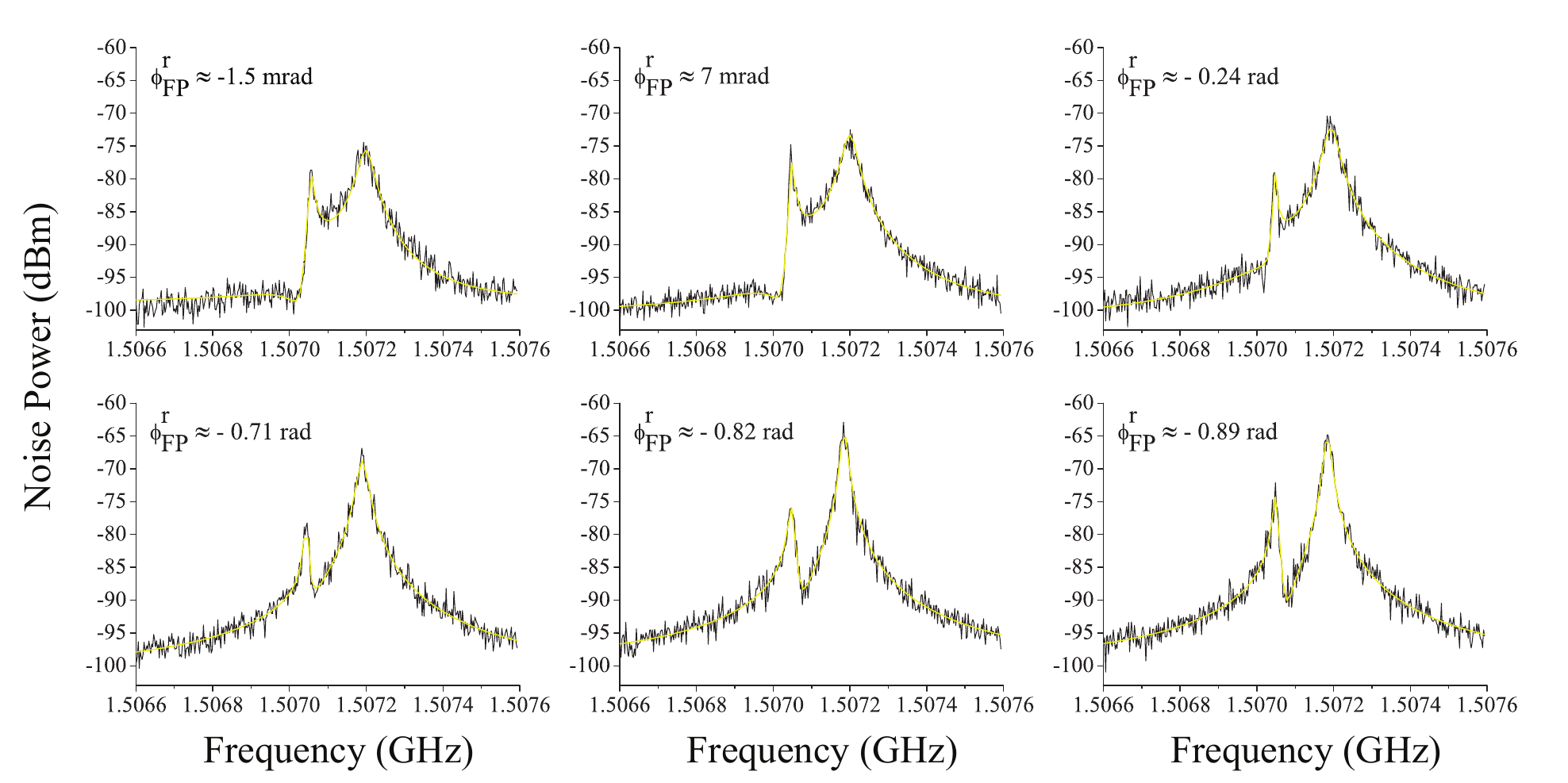}
\caption{Noise spectra recorded around the FSR $\Delta\approx$ 1.5 GHz and obtained by varying the phase shift $\phi^{\mathrm{r}}_{\mathrm{FP}}$ upon reflection on FP2. Each spectrum is fitted using $\Phi'$ instead of $\Phi$ in  Eq.\ (\ref{eq5}) and using Eq. (\ref{phase-shift-var}) with $\Phi=3.5\ \mathrm{rad}$, leading to the determination of $\phi^{\mathrm{r}}_{\mathrm{FP}}$.} \label{Fig6}
\end{figure}

In order to definitely confirm the existence of the phase locking mechanism between the non-lasing side modes, we have designed a second filtering setup in which we can change the value of $\Phi$ to observe its influence on the noise spectrum. This should allow us to completely modify the shape of the noise spectrum. To this aim, the total phase shift $\Phi$ is modified by adding a controlled phase shift to the lasing mode without disturbing the side mode phases. This is performed using reflection on another optical Fabry-Perot cavity (labeled FP2 in Fig\ \ref{set-up}), which has a FSR of 2 GHz and a finesse of 30. Using FP2 cavity in reflection configuration allows the lasing mode to undergo the cavity induced phase shift $\phi^{\mathrm{r}}_{\mathrm{FP}}$ without altering the phase of the side modes. Furthermore, this allows the side modes to be totally reflected by the cavity, thus restraining the relevant RIN signal to fall below the shot noise limit. Thanks to cavity FP2, the relative phase shift $\Phi$ is thus modified to become:
\begin{equation}
\Phi' = 2(\phi_{0}-\phi^{\mathrm{r}}_{\mathrm{FP}})-\left(\phi_{1}+\phi_{-1}\right)=\Phi-2\phi^{\mathrm{r}}_{\mathrm{FP}}\ .\label{phase-shift-var}
\end{equation}
To perform this experiment, the length of the optical cavity FP2 is servo-locked by using the reflectivity $R$ of the cavity as an error signal which we compare with a given offset. Varying this offset permits us to vary $R$ as well as $\phi^{\mathrm{r}}_{\mathrm{FP}}$. Some of the corresponding noise spectra, obtained for different values of $R$, are reproduced in Fig.\ \ref{Fig6}. Each spectrum is accompanied by its fit using  Eq.\ (\ref{eq5}). The corresponding value of $\phi^{\mathrm{r}}_{\mathrm{FP}}$ is given in each figure. We can see how a change of the relative phase between the lasing and the non-lasing modes changes the way the two beat note signals interfere. In particular, we can see that, for overall phases $\Phi'$ close to $\pi/2$, the spectrum now looks like an incoherent sum of two Lorentzians, because the interference term in Eq.\ (\ref{eq5}) vanishes.

\begin{figure}[h]
\centering
\includegraphics[width=0.5\textwidth]{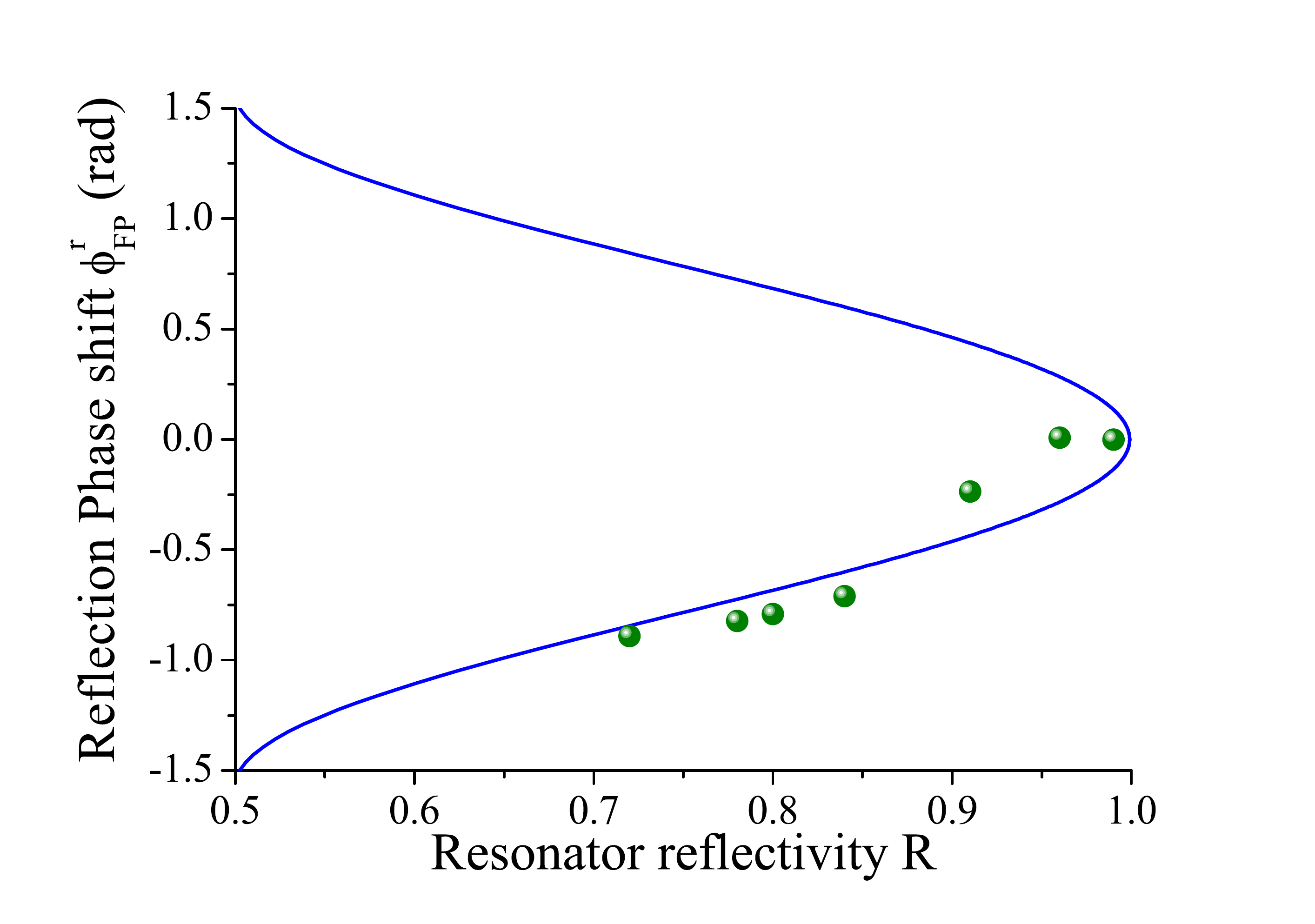}
\caption{Evolution of the phase shift $\phi^{\mathrm{r}}_{\mathrm{FP}}$ created by reflection on cavity FP2 versus reflection coefficient $R$. The full line is theoretical and the dots are extracted from fits of the experimental data like the ones of Fig.\ \ref{Fig6}.} \label{calibration}
\end{figure}
The values of $\phi^{r}_{\mathrm{FP}}$ extracted from the fits are consistent with the theoretical values of $\phi^{r}_{\mathrm{FP}}$ obtained for our cavity: Fig. \ref{calibration} shows a good agreement between theory (full line) and experiment (dots).

\begin{figure}[h]\centering
\includegraphics[width=1.0\textwidth]{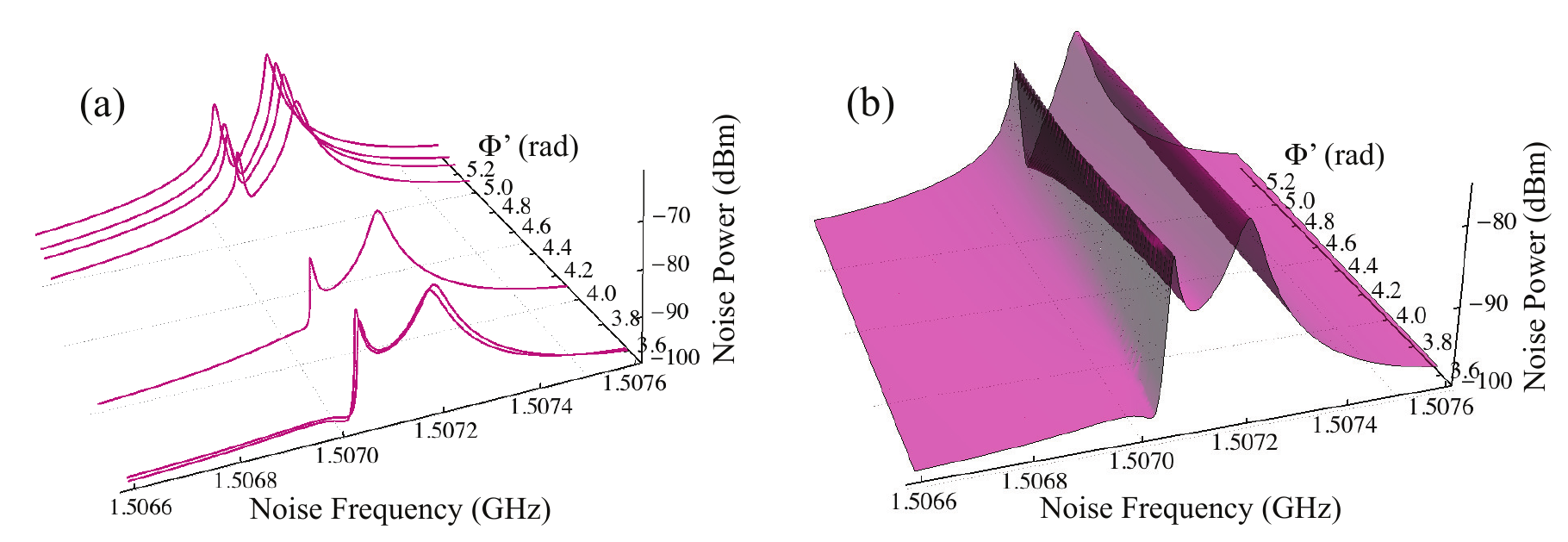}
\caption{(a) 3D representation of the fits extracted from Fig.\ \ref{Fig6}, plus an extra one which is not shown in Fig.\ \ref{Fig6}. (b) Theoretical evolution of the noise spectrum based on Eq. (\ref{eq5}). The values of the parameters used here are those extracted from the first spectrum of Fig.\ \ref{Fig6}.}
\label{comparaison-des-lolo}
\end{figure}

Finally, Fig.\ \ref{comparaison-des-lolo} reproduces the evolution of the shape of the noise spectra in the vicinity of $\Delta$ versus $\Phi'$. The left plot gathers the different fits obtained in Fig.\ \ref{Fig6} for different values of the phase. For the sake of comparison, we reproduce in the right plot the theoretical shape obtained from  Eq.\ (\ref{eq5}) by varying $\Phi$, with the following parameters: $f_{1}=1.50706\ \mathrm{GHz}$, $f_{-1}=1.5072\ \mathrm{GHz}$, $\gamma_1/2\pi =11.7\ \mathrm{kHz}$, and $\gamma_{-1}/2\pi = 46.5\ \mathrm{kHz}$. Here again, we can see that the double-peak noise spectrum varies as expected when the relative phase is varied, illustrating the fact that the three lasing and non-lasing modes corresponding to $p=-1,0,+1$ are phase locked.

\section{Conclusion}

In this paper, we have observed an unambiguous signature of phase locking between the lasing mode and the spontaneous emission noise present in the two nearest-neighbor modes of a single-frequency VECSEL. This phase locking occurs for strong enough pumping, i.e., when the modulation of the carrier density induced by the beat notes between the different modes becomes large enough to induce a significant coupling between the phases of the modes. It is worth noticing that, in the present experiment, this phase locking occurs in spite of the existence of a difference between the beat note frequencies of the lasing mode with respect to the two side modes \cite{karim}, which is absent in the case of gas lasers \cite{harris1992, loudon1994}. Moreover, we have shown that the relative phase between the three modes plays a central role in the shape of the noise spectrum.

The existence of such noise correlations between lasing and non-lasing modes opens interesting experimental as well as theoretical perspectives in the field of physics of correlations between different modes in a lasers and could be extended to the case of the different polarization modes of this kind of lasers \cite{pal2010}, and more generally, to the field of quantum optics of multimode lasers. 

\section*{Acknowledgments}
The authors thank M. P. van Exter for bringing reference \cite{VanExter1994} to their attention. They acknowledge support from the Agence National de la Recherche (project NATIF No. ANR-09-NANO-012-01), the Indo-French Center for the Promotion of Advanced Research (CEFIPRA/IFCPRA), the Triangle de la Physique, and the R\'egion Bretagne. One of the authors (VP) thanks the Council of Scientific and Industrial Research, India, for financial support.

\end{document}